\documentclass [12pt]{article}
\usepackage {graphicx}
\usepackage {longtable}

\begin{document}

\title{AN  ANALYTICAL DESCRIPTION OF SPIN EFFECTS
IN HADRON-HADRON SCATTERING VIA PMD-SQS OPTIMUM PRINCIPLE }

\maketitle

\begin{center}
D. B. Ion$^{1,2) }$, M. L. D. Ion$^{3) }$ and Adriana I. Sandru$^{1)}$
\end{center}

\begin{center}
$^{1)}$ IFIN-HH, Bucharest, P.O.Box MG-6, Magurele, Romania
\end{center}

\begin{center}
$^{2) }$TH-Division, CERN, CH-1211 Geneva 23, Switzerland
\end{center}

\begin{center}
$^{3) }$Faculty of Physics, Bucharest University, Bucharest, Romania
\end{center}

\begin{abstract}
In this paper an analytical description of spin-effects in
hadron-hadron scattering is presented by using PMD-SQS-optimum
principle in which the differential cross sections in the forward
and backward c.m. angles are considered fixed from the
experimental data. The experimental tests of the optimal
predictions, obtained by using the available phase shifts, are
discussed.
\end{abstract}

\section{Introduction}

Recently, in Ref. [1], by using reproducing kernel Hilbert space
(RKHS) methods [2-4], we described the quantum scattering of the
spinless particles by a \textit{principle of minimum distance in
the space of quantum states} (PMD-SQS). Some preliminary
experimental tests of the PMD-SQS, even in the crude form [1] when
the complications due to the particle spins are neglected, showed
that the actual experimental data for the differential cross
sections of all principal hadron-hadron [nucleon-nucleon,
antiproton-proton, meson-nucleon] scatterings at all energies
higher than 2 GeV, can be well systematized by PMD-SQS predictions
(see the papers [1]). Moreover, connections between the PMD-SQS
and the \textit{maximum entropy principle} for the statistics of
the scattering quantum channels was also recently established by
introducing quantum scattering entropies: S$_\theta$ and $S_{J }$
[5]-[7]. Then, it was shown that the experimental pion-nucleon as
well as pion-nucleus scattering entropies are well described by
optimal entropies and that the experimental data are consistent
with the principle of minimum distance in the space of quantum
states (PMD-SQS) {\}}[1]. However, the PMD-SQS in the crude form
[1] cannot describe the polarization J-spin effects.

In this paper an analytical description of spin-effects in hadron-hadron
scattering is presented by using PMD-SQS-optimum principle in which the
differential cross sections in the forward (x=+1) and backward (x=-1)
directions are considered fixed from the experimental data. An experimental
test of the optimal prediction on the logarithmic slope b is performed for
the pion-nucleon scattering at the forward c.m. angles.

\section{Optimal state description of pion-nucleon scattering}

First we present the basic definitions on the the pion-nucleon
scattering:

\begin{equation}
\pi + N \to \pi + N
\end{equation}

Therefore, let $f^{ + + }(x)$and $f^{ + - }(x)$, be the scattering
helicity amplitudes of the meson-nucleon scattering process (see
ref.[14]) written in terms of the partial helicities $f_{J - } $
sand $f_{J + } $as follows

\begin{equation}
\label{eq1}
\begin{array}{l}
 f_{ + + } \left( x \right) = \sum\limits_{J = \frac{1}{2}}^{J_{\max } }
{\left( {J + \frac{1}{2}} \right)} \left( {f_{J - } + f_{J + } }
\right)d_{\frac{1}{2}\frac{1}{2}}^J \left( x \right) \\
 f_{ + - } \left( x \right) = \sum\limits_{J = \frac{1}{2}}^{J_{\max } }
{\left( {J + \frac{1}{2}} \right)} \left( {f_{J - } - f_{J + } }
\right)d_{ - \frac{1}{2}\frac{1}{2}}^J \left( x \right) \\
 \end{array}
\end{equation}

\noindent
where the rotation functions are defined as

\begin{equation}
\label{eq2}
\begin{array}{l}
 d_{\frac{1}{2}\frac{1}{2}}^J \left( x \right) = \frac{1}{l + 1} \cdot
\left[ {\frac{1 + x}{2}} \right]^{\frac{1}{2}}\left[ {\mathop
P\limits^\bullet _{l + 1} \left( x \right) - \mathop P\limits^\bullet _l
\left( x \right)} \right] \\
 d_{ - \frac{1}{2}\frac{1}{2}}^J \left( x \right) = \frac{1}{l + 1} \cdot
\left[ {\frac{1 - x}{2}} \right]^{\frac{1}{2}}\left[ {\mathop
P\limits^\bullet _{l + 1} \left( x \right) + \mathop P\limits^\bullet _l
\left( x \right)} \right] \\
 \end{array}
\end{equation}

\noindent where $P_l (x) $are Legendre polynomials, $\mathop
P\limits^o _l (x) = \frac{d}{dx}P_l (x)$, x being the c.m.
scattering angle. The normalization of the helicity amplitudes
$f^{ + + }(x)$ and $f^{ + - }(x)$, is chosen such that the c.m.
differential cross section is given by

\begin{equation}
\label{eq3}
\frac{d\sigma }{d\Omega }\left( x \right) = \left|
{f_{ + + } \left( x \right)} \right|^2 + \left| {f_{ + - } \left(
x \right)} \right|^2
\end{equation}

Then, the elastic integrated cross section is given by

\begin{equation}
\label{eq4}
\frac{\sigma _{el} }{2\pi } = \sum {(j + \frac{1}{2}}
)\left[ {\vert f_{j + } \vert ^2 + \vert f_{j - } \vert ^2}
\right]
\end{equation}

Now, let us consider the following optimization problem:

\begin{equation}
\label{eq5} \min \left\{\sum {(j + \frac{1}{2}} )\left[ {\vert
f_{j + } \vert ^2 + \vert f_{j - } \vert ^2} \right] \right\}
\end{equation}

\noindent
when $\frac{d\sigma }{d\Omega }( + 1)\,\,\,$and $\frac{d\sigma }{d\Omega }(
- 1)$ are fixed.

We proved that the solution of this optimization problem is given by the
following results :

\begin{eqnarray}
f_{o}^{ + + }(x)= f^{ + + }(+1)\frac{K_{\frac{1}{2}\frac{1}{2}}
(x,y)}{K_{\frac{1}{2}\frac{1}{2}} ( + 1, + 1)} \\
f_{o}^{ + - }(x)= f^{ + - }(-1)\frac{K_{\frac{1}{2} - \frac{1}{2}}
(x,y)}{K_{\frac{1}{2} - \frac{1}{2}} ( - 1, - 1)}
\end{eqnarray}

\noindent where the functions K(x,y) are the reproducing kernels
[2] expressed in terms of the rotation function by

\begin{eqnarray}
K_{\frac{1}{2}\frac{1}{2}} (x,y) = \sum\limits_{1 / 2}^{J_o } ( j
+ \frac{1}{2})d_{\frac{1}{2}\frac{1}{2}}^j
(x)d_{\frac{1}{2}\frac{1}{2}}^j (y) \\
K_{\frac{1}{2} - \frac{1}{2}} (x,y) = \sum\limits_{1 / 2}^{J_o } (
j + \frac{1}{2})d_{\frac{1}{2} - \frac{1}{2}}^j (x)d_{\frac{1}{2}
- \frac{1}{2}}^j (y)
\end{eqnarray}

\noindent
while the optimal angular momentum is given by

\begin{equation}
\label{jopt} J_0 = \sqrt {\frac{4\pi }{\sigma _{el} }\left[
{\frac{d\sigma }{d\Omega }( + 1) + \frac{d\sigma }{d\Omega }( -
1)} \right] + \frac{1}{4}} - 1
\end{equation}

Now, let us consider the logarithmic slope b of the forward diffraction peak
defined by

\begin{eqnarray}
b=\frac{d}{dt}\left[ {\ln \frac{d\sigma }{dt}(s,t)} \right]_{t =
0}
\end{eqnarray}

Then, using the definition of the rotation functions, from
(70-(\ref{eq5}) we obtain the following optimal slope $b_{o}$:

\begin{eqnarray}
b_o = \frac{\lambda ^2}{4}\left[ {\frac{4\pi }{\sigma _{el}
}\left( {\frac{d\sigma }{d\Omega }( + 1) + \frac{d\sigma }{d\Omega
}( - 1)} \right) - 1} \right]
\end{eqnarray}

Optimal predictions on the differential cross section $\frac{d\sigma _o
}{d\Omega }(x)$ and also for the spin-polarization parameters (P$_{o }$,
R$_{o }$, A$_{o})$, are as follows.$_{ }$

\begin{eqnarray}
\label{eq6} \frac{d\sigma _o }{d\Omega }(x) = \frac{d\sigma
}{d\Omega }( + 1)\left[ {\frac{K_{\frac{1}{2}\frac{1}{2}} (x, +
1)}{K_{\frac{1}{2}\frac{1}{2}} ( + 1, + 1)}} \right]^2 +
\frac{d\sigma }{d\Omega }( - 1)\left[ {\frac{K_{ -
\frac{1}{2}\frac{1}{2}} (x, - 1)}{K_{ - \frac{1}{2}\frac{1}{2}} (
- 1, - 1)}} \right]^2 \\
P_o \frac{d\sigma _o }{d\Omega }(x) = 2\sqrt {\frac{d\sigma
}{d\Omega }( + 1)} \sqrt {\frac{d\sigma }{d\Omega }( - 1)} \left[
{\frac{K_{\frac{1}{2}\frac{1}{2}} (x, +
1)}{K_{\frac{1}{2}\frac{1}{2}} ( + 1, + 1)}} \right]\left[
{\frac{K_{ - \frac{1}{2}\frac{1}{2}} (x, - 1)}{K_{ -
\frac{1}{2}\frac{1}{2}} ( - 1, - 1)}} \right]\sin \phi (x)
\\
R_o \frac{d\sigma _o }{d\Omega }(x) = 2\sqrt {\frac{d\sigma
}{d\Omega }( + 1)} \sqrt {\frac{d\sigma }{d\Omega }( - 1)} \left[
{\frac{K_{\frac{1}{2}\frac{1}{2}} (x, +
1)}{K_{\frac{1}{2}\frac{1}{2}} ( + 1, + 1)}} \right]\left[
{\frac{K_{ - \frac{1}{2}\frac{1}{2}} (x, - 1)}{K_{ -
\frac{1}{2}\frac{1}{2}} ( - 1, - 1)}} \right]\cos \phi (x)
\\
A_o \frac{d\sigma _o }{d\Omega }(x) = \frac{d\sigma }{d\Omega }( +
1)\left[ {\frac{K_{\frac{1}{2}\frac{1}{2}} (x, +
1)}{K_{\frac{1}{2}\frac{1}{2}} ( + 1, + 1)}} \right]^2 -
\frac{d\sigma }{d\Omega }( - 1)\left[ {\frac{K_{ -
\frac{1}{2}\frac{1}{2}} (x, - 1)}{K_{ - \frac{1}{2}\frac{1}{2}} (
- 1, - 1)}} \right]^2
\end{eqnarray}

 where

\begin{equation}
\label{eq8}
\cos \phi (x) = \frac{Re\{[f^{ + + }( + 1)]^\ast f^{ + - }( - 1)\}}{\vert
f^{ + + }( + 1)\vert \vert f^{ + - }( - 1)\vert },
\quad
\sin \phi (x) = \frac{Im\{[f^{ + + }( + 1)]^\ast f^{ + - }( - 1)\}}{\vert
f^{ + + }( + 1)\vert \vert f^{ + - }( - 1)\vert }
\end{equation}

Finally, we note that in ref. [15] we proved the following optimal
inequality

\begin{equation}
\label{eq9}
b_o = \frac{\lambda ^2}{4}\left[ {\frac{4\pi }{\sigma _{el} }\left(
{\frac{d\sigma }{d\Omega }( + 1) + \frac{d\sigma }{d\Omega }( - 1)} \right)
- 1} \right] \le b_{\exp }
\end{equation}

\noindent which includes in a more general and exact form the
unitarity bounds derived by Martin [8] and Martin-MacDowell [9]
(see also ref.[10]) and Ion [1],[11].

\section{Experimental test of the PMD-SQS-optimal predictions}

For an experimental test of the optimal result (13) the numerical
values of the slopes b$_{o }$ and b$_{exp }$ are calculated
directly by reconstruction of the helicity amplitudes from the
experimental phase shifts (EPS) solutions of Holer et al. [14].
These results are given in the Tables 1-2 and are displayed in Fig
1-2.

\section{Conclusions}

The main results and conclusions obtained in this paper can be summarized as
follows:

(i) The optimal state dominance in hadron-hadron scattering at
small transfer momenta for $p_{LAB}\geq 2$  GeV/c is a fact well
evidenced experimentally by the results presented in Figs. 1-2.

(ii) In the low energy region, the optimal slope (13) is in good agreement
with the experimental data at discrete values of energy between the
resonances positions or/and in the region corresponding to the diffractive
resonances see Figs. 1-2.

Finally, we hope that our results as well as those from ref. [15]
are new steps towards an analytic description of the quantum
scattering in terms of an optimum principle, namely, the
\textit{principle of minimum distance in space of quantum state}
(PMD-SQS) [1].

\begin{longtable}[p]
{|p{36pt}|p{37pt}|p{33pt}|p{33pt}|p{33pt}|p{30pt}|p{36pt}|p{37pt}|p{43pt}|}
\caption{PMD-SQS optimal predictions for $\pi ^ + P \to \pi ^ +
P$, calculated using the experimental phase shifts solutions from
ref. [14]} \label{tab1}
\\a & a & a & a & a & a & a & a & a  \kill
\hline P$_{LAB}$& (1/p$_{cm})^{2}$& $\frac{d\sigma }{d\Omega }( +
1)$& $\frac{d\sigma }{d\Omega }( - 1)$& $\sigma _{el} $& J$_{0}$&
b$_{o}$& b$_{o + 1}$&
b$_{exp}$ \\
(GeV/c)& (GeV$^{ - 2})$& (mb/sr)& (mb/sr)& (mb)& & (GeV$^{ - 2})$&
(GeV$^{ - 2})$&
(GeV$^{ - 3})$ \\
\hline
\endfirsthead

\multicolumn{9}{c}%
{{\bfseries \tablename\ \thetable{} -- continued from previous page}} \\
\hline P$_{LAB}$& (1/p$_{cm})^{2}$& $\frac{d\sigma }{d\Omega }( +
1)$& $\frac{d\sigma }{d\Omega }( - 1)$& $\sigma _{el} $& J$_{0}$&
b$_{o}$& b$_{o + 1}$&
b$_{exp}$ \\
(GeV/c)& (GeV$^{ - 2})$& (mb/sr)& (mb/sr)& (mb)& & (GeV$^{ - 2})$&
(GeV$^{ - 2})$&
(GeV$^{ - 3})$ \\
\hline
\endhead

0.020& 3305.477& 0.185& 0.185& 2.329& 0.500& 826.369& 0.000&
0.000 \\
\hline 0.040& 831.948& 0.117& 0.273& 2.426& 0.507& 212.175&
-81.659&
-255.100 \\
\hline 0.060& 373.735& 0.042& 0.473& 2.955& 0.562& 111.242&
-76.667&
-655.969 \\
\hline 0.079& 218.466& 0.003& 0.834& 4.255& 0.650& 80.360&
-54.150&
-6650.37 \\
\hline 0.097& 147.078& 0.027& 1.254& 6.057& 0.705& 60.956&
-34.742&
-382.911 \\
\hline 0.112& 111.845& 0.133& 1.831& 8.638& 0.763& 51.949&
-22.531&
14.998 \\
\hline 0.130& 84.501& 0.425& 2.911& 13.489& 0.832& 44.518&
-12.769&
46.680 \\
\hline 0.153& 62.493& 1.294& 4.971& 24.018& 0.878& 35.590& -5.047&
45.665 \\
\hline 0.172& 50.486& 2.621& 7.570& 37.559& 0.913& 30.414& -1.554&
36.764 \\
\hline 0.185& 44.276& 4.091& 9.466& 49.167& 0.927& 27.285& 0.505&
31.826 \\
\hline 0.200& 38.530& 6.621& 12.743& 68.315& 0.952& 24.677& 2.099&
28.020 \\
\hline 0.218& 33.100& 11.275& 18.196& 101.316& 0.976& 21.973&
3.297&
24.066 \\
\hline 0.247& 26.651& 22.300& 28.265& 169.036& 1.002& 18.383&
4.383&
19.386 \\
\hline 0.267& 23.333& 30.173& 32.196& 203.762& 1.024& 16.604&
5.021&
16.972 \\
\hline 0.280& 21.531& 32.650& 31.802& 207.611& 1.037& 15.616&
5.255&
15.690 \\
\hline 0.290& 20.299& 33.087& 30.510& 202.992& 1.046& 14.905&
5.320&
14.776 \\
\hline 0.295& 19.727& 33.442& 29.397& 199.639& 1.051& 14.575&
5.450&
14.316 \\
\hline 0.301& 19.076& 33.287& 27.945& 193.442& 1.056& 14.201&
5.543&
13.785 \\
\hline 0.305& 18.662& 32.962& 26.816& 188.205& 1.059& 13.956&
5.603&
13.460 \\
\hline 0.310& 18.166& 32.487& 25.413& 180.947& 1.067& 13.720&
5.705&
13.111 \\
\hline 0.320& 17.238& 31.198& 22.436& 165.767& 1.077& 13.213&
5.883&
12.391 \\
\hline 0.331& 16.308& 28.798& 19.733& 150.463& 1.074& 12.448&
5.729&
11.601 \\
\hline 0.351& 14.822& 25.715& 14.507& 122.777& 1.090& 11.550&
6.047&
10.492 \\
\hline 0.378& 13.156& 20.937& 10.134& 92.640& 1.113& 10.573&
6.052&
9.381 \\
\hline 0.408& 11.654& 16.983& 7.542& 71.111& 1.141& 9.713& 5.830&
8.490 \\
\hline 0.427& 10.850& 15.042& 5.519& 60.807& 1.121& 8.813& 5.720&
7.679 \\
\hline 0.456& 9.797& 12.345& 3.528& 47.050& 1.119& 7.934& 5.626&
6.381 \\
\hline 0.490& 8.774& 10.059& 2.446& 37.329& 1.112& 7.041& 5.234&
5.382 \\
\hline 0.532& 7.748& 8.407& 1.372& 28.181& 1.147& 6.510& 5.325&
4.729 \\
\hline 0.573& 6.936& 6.766& 0.700& 21.827& 1.133& 5.720& 5.020&
4.864 \\
\hline 0.614& 6.266& 5.409& 0.233& 17.710& 1.062& 4.706& 4.447&
4.805 \\
\hline 0.658& 5.668& 4.386& 0.089& 14.446& 1.035& 4.099& 3.989&
5.937 \\
\hline 0.675& 5.463& 3.896& 0.082& 13.190& 1.010& 3.810& 3.704&
5.662 \\
\hline 0.705& 5.134& 3.358& 0.057& 11.125& 1.027& 3.667& 3.585&
7.180 \\
\hline 0.725& 4.934& 2.959& 0.037& 10.166& 0.988& 3.335& 3.279&
7.121 \\
\hline 0.750& 4.704& 2.523& 0.021& 8.767& 0.974& 3.112& 3.076&
7.092 \\
\hline 0.777& 4.476& 2.728& 0.081& 11.235& 0.842& 2.397& 2.295&
4.894 \\
\hline 0.800& 4.298& 1.981& 0.003& 7.050& 0.946& 2.726& 2.720&
2.383 \\
\hline 0.822& 4.139& 1.851& 0.002& 6.467& 0.962& 2.691& 2.687&
5.419 \\
\hline 0.851& 3.946& 1.868& 0.016& 6.383& 0.989& 2.671& 2.640&
3.371 \\
\hline 0.875& 3.798& 2.035& 0.060& 6.818& 1.028& 2.719& 2.613&
4.051 \\
\hline 0.895& 3.683& 2.183& 0.159& 7.683& 1.020& 2.606& 2.367&
3.029 \\
\hline 0.923& 3.532& 2.458& 0.315& 8.999& 1.030& 2.536& 2.147&
2.115 \\
\hline 0.954& 3.378& 2.812& 0.505& 10.365& 1.067& 2.551& 2.034&
2.057 \\
\hline 0.975& 3.281& 3.012& 0.618& 11.061& 1.091& 2.562& 1.986&
2.634 \\
\hline 1.000& 3.171& 3.507& 0.754& 11.758& 1.192& 2.818& 2.179&
1.543 \\
\hline 1.030& 3.049& 3.579& 0.986& 11.731& 1.267& 2.965& 2.160&
2.684 \\
\hline 1.055& 2.954& 3.687& 1.034& 11.903& 1.288& 2.942& 2.136&
3.021 \\
\hline 1.080& 2.864& 3.816& 1.093& 12.011& 1.321& 2.961& 2.142&
2.310 \\
\hline 1.113& 2.753& 4.249& 1.326& 12.298& 1.439& 3.233& 2.300&
4.100 \\
\hline 1.154& 2.625& 4.902& 1.594& 12.951& 1.560& 3.481& 2.466&
5.165 \\
\hline 1.174& 2.567& 5.248& 1.714& 13.390& 1.605& 3.552& 2.520&
6.220 \\
\hline 1.210& 2.469& 6.112& 2.065& 14.198& 1.736& 3.850& 2.722&
6.513 \\
\hline 1.235& 2.405& 6.754& 2.239& 14.778& 1.810& 3.996& 2.851&
6.239 \\
\hline 1.280& 2.297& 8.371& 2.494& 15.826& 1.979& 4.379& 3.242&
6.929 \\
\hline 1.324& 2.200& 10.306& 2.774& 16.680& 2.179& 4.869& 3.720&
7.720 \\
\hline 1.360& 2.126& 11.929& 2.728& 17.133& 2.317& 5.182& 4.119&
8.469 \\
\hline 1.400& 2.049& 13.468& 2.608& 17.546& 2.430& 5.387& 4.430&
8.160 \\
\hline 1.430& 1.995& 14.381& 2.487& 17.605& 2.506& 5.508& 4.622&
8.869 \\
\hline 1.473& 1.923& 15.282& 2.189& 17.401& 2.587& 5.584& 4.824&
8.308 \\
\hline 1.505& 1.872& 15.777& 1.912& 17.265& 2.623& 5.557& 4.906&
7.235 \\
\hline 1.550& 1.804& 15.023& 1.591& 16.169& 2.628& 5.373& 4.816&
7.097 \\
\hline 1.590& 1.748& 14.457& 1.236& 15.287& 2.626& 5.201& 4.757&
6.626 \\
\hline 1.640& 1.683& 13.875& 0.879& 14.218& 2.646& 5.065& 4.738&
6.513 \\
\hline 1.680& 1.634& 13.428& 0.602& 13.232& 2.684& 5.033& 4.799&
6.425 \\
\hline 1.720& 1.587& 13.034& 0.450& 12.729& 2.683& 4.885& 4.709&
6.383 \\
\hline 1.760& 1.543& 12.617& 0.324& 12.065& 2.705& 4.814& 4.684&
6.210 \\
\hline 1.800& 1.502& 12.279& 0.241& 11.602& 2.716& 4.715& 4.617&
5.858 \\
\hline 1.840& 1.462& 11.843& 0.211& 11.032& 2.739& 4.653& 4.566&
5.896 \\
\hline 1.880& 1.425& 11.883& 0.135& 10.757& 2.780& 4.644& 4.587&
5.958 \\
\hline 1.920& 1.389& 11.822& 0.098& 10.320& 2.843& 4.693& 4.651&
5.883 \\
\hline 1.980& 1.338& 11.836& 0.054& 9.958& 2.906& 4.686& 4.663&
5.865 \\
\hline 2.03& 1.299& 11.976& 0.043& 9.603& 2.997& 4.783& 4.765&
6.044 \\
\hline 2.07& 1.269& 12.061& 0.037& 9.465& 3.039& 4.779& 4.763&
5.881 \\
\hline 2.15& 1.213& 12.500& 0.054& 9.179& 3.176& 4.910& 4.887&
5.919 \\
\hline 2.20& 1.181& 12.973& 0.086& 9.075& 3.282& 5.043& 5.007&
5.997 \\
\hline 2.28& 1.132& 13.706& 0.154& 8.894& 3.453& 5.258& 5.197&
6.228 \\
\hline 2.34& 1.098& 14.547& 0.187& 8.829& 3.607& 5.482& 5.409&
6.387 \\
\hline 2.46& 1.036& 15.840& 0.273& 8.702& 3.849& 5.765& 5.663&
6.485 \\
\hline 2.56& 0.989& 16.725& 0.274& 8.585& 4.013& 5.903& 5.804&
6.458 \\
\hline 2.75& 0.910& 18.072& 0.239& 8.328& 4.280& 6.059& 5.977&
6.479 \\
\hline 3.00& 0.824& 19.115& 0.167& 7.803& 4.595& 6.190& 6.135&
6.848 \\
\hline 3.40& 0.715& 20.284& 0.093& 7.098& 5.027& 6.269& 6.240&
6.942 \\
\hline 3.65& 0.660& 21.180& 0.053& 6.886& 5.245& 6.231& 6.215&
6.708 \\
\hline 4.00& 0.596& 23.088& 0.062& 6.590& 5.663& 6.432& 6.415&
7.018 \\
\hline 5.00& 0.467& 26.653& 0.042& 5.751& 6.654& 6.690& 6.680&
6.989 \\
\hline 6.00& 0.383& 30.776& 0.031& 5.259& 7.594& 6.957& 6.950&
7.302 \\
\hline 10.00& 0.223& 44.885& 0.097& 4.414& 10.328& 7.087& 7.072&
7.369 \\
\hline
\end{longtable}

\begin{longtable}[p]
{|p{36pt}|p{37pt}|p{33pt}|p{33pt}|p{33pt}|p{30pt}|p{36pt}|p{37pt}|p{40pt}|}
\caption{PMD-SQS optimal predictions for $\pi ^ + P \to \pi ^ +
P$, calculated using the experimental phase shifts solutions from
ref. [14]} \label{tab2}
\\a & a & a & a & a & a & a & a & a  \kill
\hline P$_{LAB}$& (1/p$_{cm})^{2}$& $\frac{d\sigma }{d\Omega }( +
1)$& $\frac{d\sigma }{d\Omega }( - 1)$& $\sigma _{el} $& J$_{0}$&
b$_{o}$& b$_{o + 1}$&
b$_{exp}$ \\
(GeV/c)& (GeV$^{ - 2})$& (mb/sr)& (mb/sr)& (mb)& & (GeV$^{ - 2})$&
(GeV$^{ - 2})$&
(GeV$^{ - 3})$ \\
\hline
\endfirsthead

\multicolumn{9}{c}%
{{\bfseries \tablename\ \thetable{} -- continued from previous page}} \\
\hline P$_{LAB}$& (1/p$_{cm})^{2}$& $\frac{d\sigma }{d\Omega }( +
1)$& $\frac{d\sigma }{d\Omega }( - 1)$& $\sigma _{el} $& J$_{0}$&
b$_{o}$& b$_{o + 1}$&
b$_{exp}$ \\
(GeV/c)& (GeV$^{ - 2})$& (mb/sr)& (mb/sr)& (mb)& & (GeV$^{ - 2})$&
(GeV$^{ - 2})$&
(GeV$^{ - 3})$ \\
\hline
\endhead

0.020& 3305.477& 0.146& 0.146& 1.840& 0.500& 826.369& 0.000&
0.000 \\
\hline 0.04o& 831.948& 0.135& 0.144& 1.755& 0.498& 206.915&
-7.183&
-17.101 \\
\hline 0.060& 373.735& 0.147& 0.124& 1.739& 0.486& 89.461& 5.960&
3.735 \\
\hline 0.079& 218.466& 0.214& 0.099& 2.024& 0.481& 51.553& 17.938&
22.395 \\
\hline 0.097& 147.078& 0.240& 0.072& 2.076& 0.463& 32.714& 16.711&
17.401 \\
\hline 0.112& 111.845& 0.277& 0.038& 2.095& 0.461& 24.770& 18.437&
18.375 \\
\hline 0.130& 84.501& 0.360& 0.007& 2.247& 0.519& 22.326& 21.460&
25.665 \\
\hline 0.153& 62.493& 0.510& 0.011& 2.853& 0.596& 20.242& 19.456&
24.444 \\
\hline 0.172& 50.486& 0.766& 0.070& 3.982& 0.699& 20.666& 17.890&
24.759 \\
\hline 0.185& 44.276& 0.984& 0.168& 5.009& 0.772& 20.934& 16.255&
24.908 \\
\hline 0.200& 38.530& 1.371& 0.390& 6.860& 0.864& 21.439& 14.564&
25.689 \\
\hline 0.218& 33.100& 1.985& 0.867& 10.142& 0.945& 20.963& 12.078&
25.069 \\
\hline 0.247& 26.651& 3.357& 1.867& 17.184& 1.017& 18.788& 9.693&
20.694 \\
\hline 0.267& 23.333& 4.112& 2.667& 21.502& 1.052& 17.276& 8.184&
17.930 \\
\hline 0.280& 21.531& 4.144& 2.977& 22.281& 1.066& 16.236& 7.198&
16.414 \\
\hline 0.290& 20.299& 4.054& 3.079& 22.208& 1.070& 15.409& 6.568&
15.127 \\
\hline 0.295& 19.727& 3.956& 3.128& 21.966& 1.074& 15.055& 6.231&
14.511 \\
\hline 0.301& 19.076& 3.842& 3.140& 21.569& 1.078& 14.632& 5.906&
13.870 \\
\hline 0.305& 18.662& 3.716& 3.100& 21.084& 1.077& 14.288& 5.668&
13.312 \\
\hline 0.310& 18.166& 3.526& 3.029& 20.248& 1.078& 13.932& 5.396&
12.668 \\
\hline 0.320& 17.238& 3.251& 2.900& 19.093& 1.073& 13.136& 4.911&
11.270 \\
\hline 0.331& 16.308& 2.900& 2.696& 17.726& 1.054& 12.096& 4.305&
9.950 \\
\hline 0.351& 14.822& 2.314& 2.301& 15.157& 1.019& 10.474& 3.403&
7.325 \\
\hline 0.378& 13.156& 1.651& 1.806& 12.043& 0.964& 8.574& 2.377&
3.563 \\
\hline 0.408& 11.654& 1.210& 1.483& 10.323& 0.879& 6.640& 1.380&
-0.334 \\
\hline 0.427& 10.850& 1.079& 1.050& 10.050& 0.706& 4.507& 0.946&
-3.973 \\
\hline 0.456& 9.797& 1.185& 1.067& 9.564& 0.791& 4.798& 1.364&
-0.107 \\
\hline 0.490& 8.774& 1.637& 1.071& 10.688& 0.853& 4.790& 2.028&
2.468 \\
\hline 0.532& 7.748& 1.866& 1.246& 10.636& 0.982& 5.185& 2.334&
2.193 \\
\hline 0.573& 6.936& 2.495& 1.379& 11.918& 1.082& 5.349& 2.829&
4.136 \\
\hline 0.614& 6.266& 3.341& 1.320& 13.260& 1.160& 5.354& 3.394&
6.863 \\
\hline 0.658& 5.668& 5.251& 1.516& 15.402& 1.402& 6.406& 4.654&
11.622 \\
\hline 0.675& 5.463& 5.869& 1.585& 16.407& 1.441& 6.431& 4.774&
10.935 \\
\hline 0.705& 5.134& 7.139& 1.566& 17.806& 1.529& 6.602& 5.183&
10.529 \\
\hline 0.725& 4.934& 7.238& 1.388& 17.790& 1.519& 6.283& 5.073&
10.413 \\
\hline 0.750& 4.704& 6.794& 1.041& 16.931& 1.463& 5.663& 4.754&
8.870 \\
\hline 0.777& 4.476& 5.950& 0.719& 14.958& 1.419& 5.151& 4.475&
9.318 \\
\hline 0.800& 4.298& 5.567& 0.424& 13.119& 1.447& 5.092& 4.655&
11.247 \\
\hline 0.822& 4.139& 5.332& 0.296& 12.151& 1.464& 4.989& 4.671&
12.973 \\
\hline 0.851& 3.946& 5.873& 0.189& 12.277& 1.541& 5.135& 4.944&
14.188 \\
\hline 0.875& 3.798& 7.117& 0.151& 13.584& 1.641& 5.435& 5.303&
14.491 \\
\hline 0.895& 3.683& 8.524& 0.131& 15.314& 1.712& 5.619& 5.520&
13.478 \\
\hline 0.923& 3.532& 11.592& 0.135& 18.379& 1.875& 6.197& 6.116&
13.609 \\
\hline 0.954& 3.378& 15.096& 0.189& 21.664& 2.019& 6.643& 6.550&
13.139 \\
\hline 0.975& 3.281& 17.059& 0.237& 23.422& 2.087& 6.790& 6.686&
13.470 \\
\hline 1.000& 3.171& 19.475& 0.253& 25.402& 2.164& 6.945& 6.846&
11.733 \\
\hline 1.030& 3.049& 18.243& 0.350& 24.370& 2.137& 6.546& 6.408&
11.839 \\
\hline 1.055& 2.954& 16.727& 0.203& 22.918& 2.088& 6.116& 6.034&
10.378 \\
\hline 1.080& 2.864& 14.490& 0.177& 20.993& 2.005& 5.569& 5.494&
8.474 \\
\hline 1.113& 2.753& 12.188& 0.105& 18.286& 1.949& 5.126& 5.076&
7.380 \\
\hline 1.154& 2.625& 10.354& 0.063& 15.651& 1.935& 4.833& 4.800&
6.605 \\
\hline 1.174& 2.567& 9.985& 0.085& 14.774& 1.969& 4.856& 4.810&
7.245 \\
\hline 1.210& 2.469& 8.142& 0.072& 11.935& 1.983& 4.721& 4.674&
8.381 \\
\hline 1.235& 2.405& 9.990& 0.131& 13.495& 2.110& 5.064& 4.991&
7.676 \\
\hline 1.280& 2.297& 10.063& 0.188& 12.973& 2.190& 5.127& 5.022&
7.619 \\
\hline 1.324& 2.200& 10.496& 0.258& 12.483& 2.328& 5.403& 5.261&
7.343 \\
\hline 1.360& 2.126& 10.588& 0.280& 11.947& 2.418& 5.544& 5.388&
8.192 \\
\hline 1.400& 2.049& 10.944& 0.304& 11.481& 2.544& 5.795& 5.625&
8.576 \\
\hline 1.430& 1.995& 11.214& 0.369& 11.298& 2.624& 5.928& 5.724&
9.010 \\
\hline 1.473& 1.923& 11.226& 0.309& 10.762& 2.704& 5.993& 5.820&
8.693 \\
\hline 1.505& 1.872& 11.443& 0.308& 10.893& 2.716& 5.876& 5.709&
7.948 \\
\hline 1.550& 1.804& 11.496& 0.298& 10.650& 2.764& 5.826& 5.667&
7.940 \\
\hline 1.590& 1.748& 10.504& 0.443& 10.029& 2.737& 5.558& 5.315&
8.458 \\
\hline 1.640& 1.683& 11.858& 0.206& 10.190& 2.889& 5.838& 5.731&
8.038 \\
\hline 1.680& 1.634& 12.158& 0.177& 9.899& 2.988& 5.986& 5.894&
8.250 \\
\hline 1.720& 1.587& 12.398& 0.141& 9.833& 3.034& 5.962& 5.890&
8.025 \\
\hline 1.760& 1.543& 12.827& 0.127& 9.688& 3.129& 6.097& 6.033&
8.204 \\
\hline 1.800& 1.502& 13.270& 0.102& 9.649& 3.203& 6.163& 6.113&
8.103 \\
\hline 1.840& 1.462& 13.716& 0.020& 9.393& 3.316& 6.351& 6.341&
8.462 \\
\hline 1.880& 1.425& 14.604& 0.069& 9.731& 3.382& 6.393& 6.361&
8.368 \\
\hline 1.920& 1.389& 15.142& 0.049& 9.738& 3.456& 6.459& 6.437&
8.194 \\
\hline 1.980& 1.338& 16.207& 0.027& 9.768& 3.597& 6.654& 6.642&
8.299 \\
\hline 2.030& 1.299& 16.747& 0.014& 9.641& 3.701& 6.770& 6.764&
8.233 \\
\hline 2.070& 1.269& 17.193& 0.006& 9.587& 3.774& 6.836& 6.834&
8.181 \\
\hline 2.150& 1.213& 17.646& 0.000& 9.328& 3.901& 6.907& 6.907&
7.983 \\
\hline 2.200& 1.181& 18.039& 0.003& 9.112& 4.013& 7.050& 7.048&
8.226 \\
\hline 2.280& 1.132& 18.320& 0.005& 8.908& 4.109& 7.032& 7.030&
8.119 \\
\hline 2.340& 1.098& 18.611& 0.011& 8.732& 4.201& 7.081& 7.077&
8.103 \\
\hline 2.460& 1.036& 18.712& 0.021& 8.401& 4.317& 6.996& 6.988&
7.853 \\
\hline 2.560& 0.989& 18.970& 0.031& 8.233& 4.409& 6.921& 6.910&
7.782 \\
\hline 2.750& 0.910& 19.759& 0.033& 7.824& 4.660& 7.006& 6.994&
7.648 \\
\hline 3.000& 0.824& 21.451& 0.025& 7.593& 4.983& 7.115& 7.106&
7.919 \\
\hline 3.400& 0.715& 23.638& 0.014& 7.095& 5.492& 7.309& 7.305&
8.036 \\
\hline 3.650& 0.660& 24.337& 0.008& 6.800& 5.726& 7.262& 7.259&
8.097 \\
\hline 4.000& 0.596& 25.774& 0.014& 6.544& 6.055& 7.234& 7.230&
7.829 \\
\hline 5.000& 0.467& 30.771& 0.006& 5.818& 7.169& 7.641& 7.639&
8.281 \\
\hline 6.000& 0.383& 34.735& 0.028& 5.329& 8.068& 7.759& 7.753&
8.507 \\
\hline 10.000& 0.223& 52.147& 0.019& 4.601& 10.947& 7.891& 7.888&
8.418 \\
\hline
\end{longtable}

\begin{figure}[p]
\centerline{\includegraphics[width=16cm]{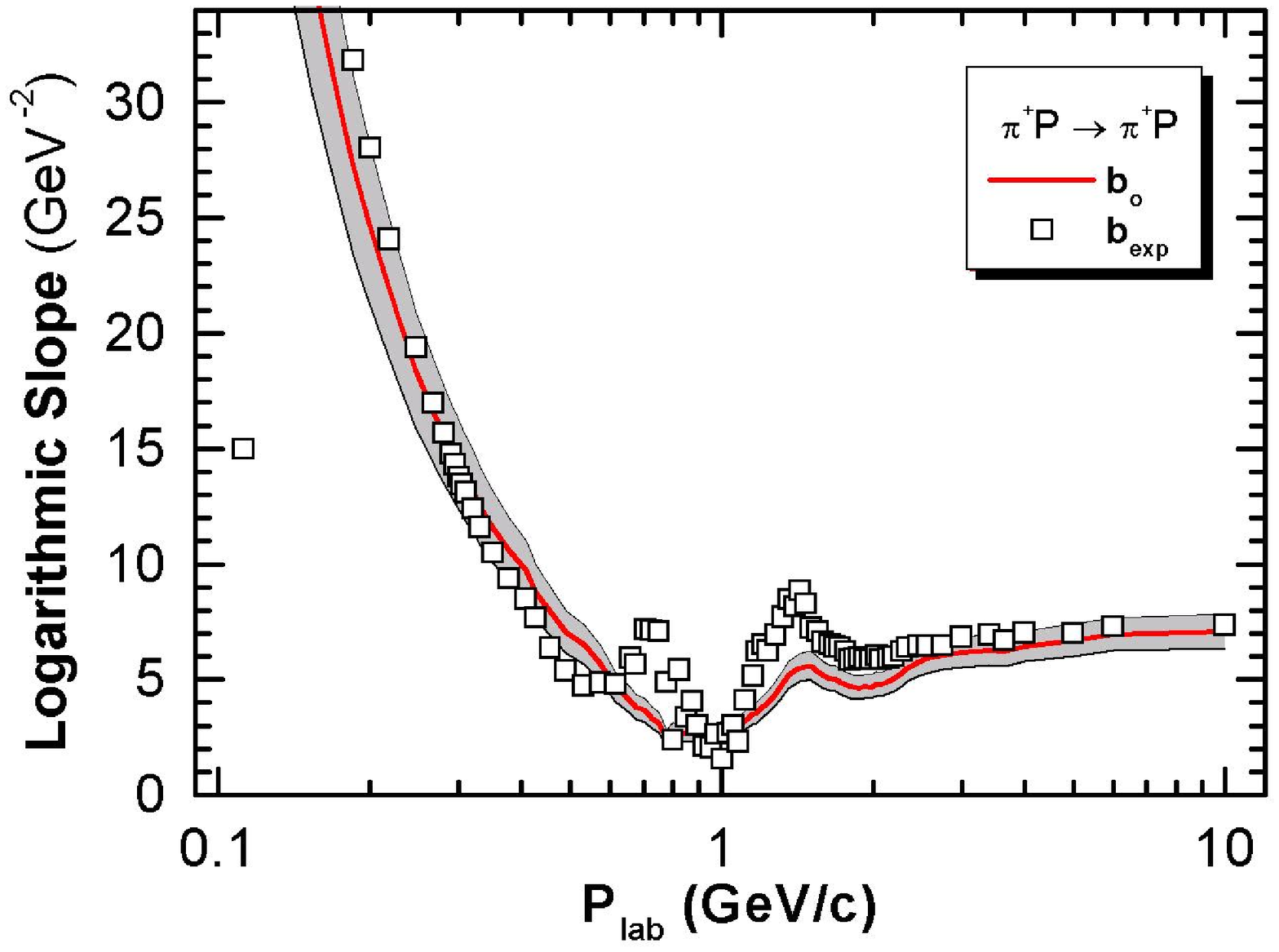}}
\caption{The experimental logarithmic slopes (b$_{exp})$ of the
diffraction peak, for the forward $\pi ^ + P \to \pi ^ + P$
scattering, are compared with the PMD-SQS-optimal predictions
b$_{o}$ (13) (see the text and Table 1). } \label{fig1}
\end{figure}

\begin{figure}[p]
\centerline{\includegraphics[width=16cm]{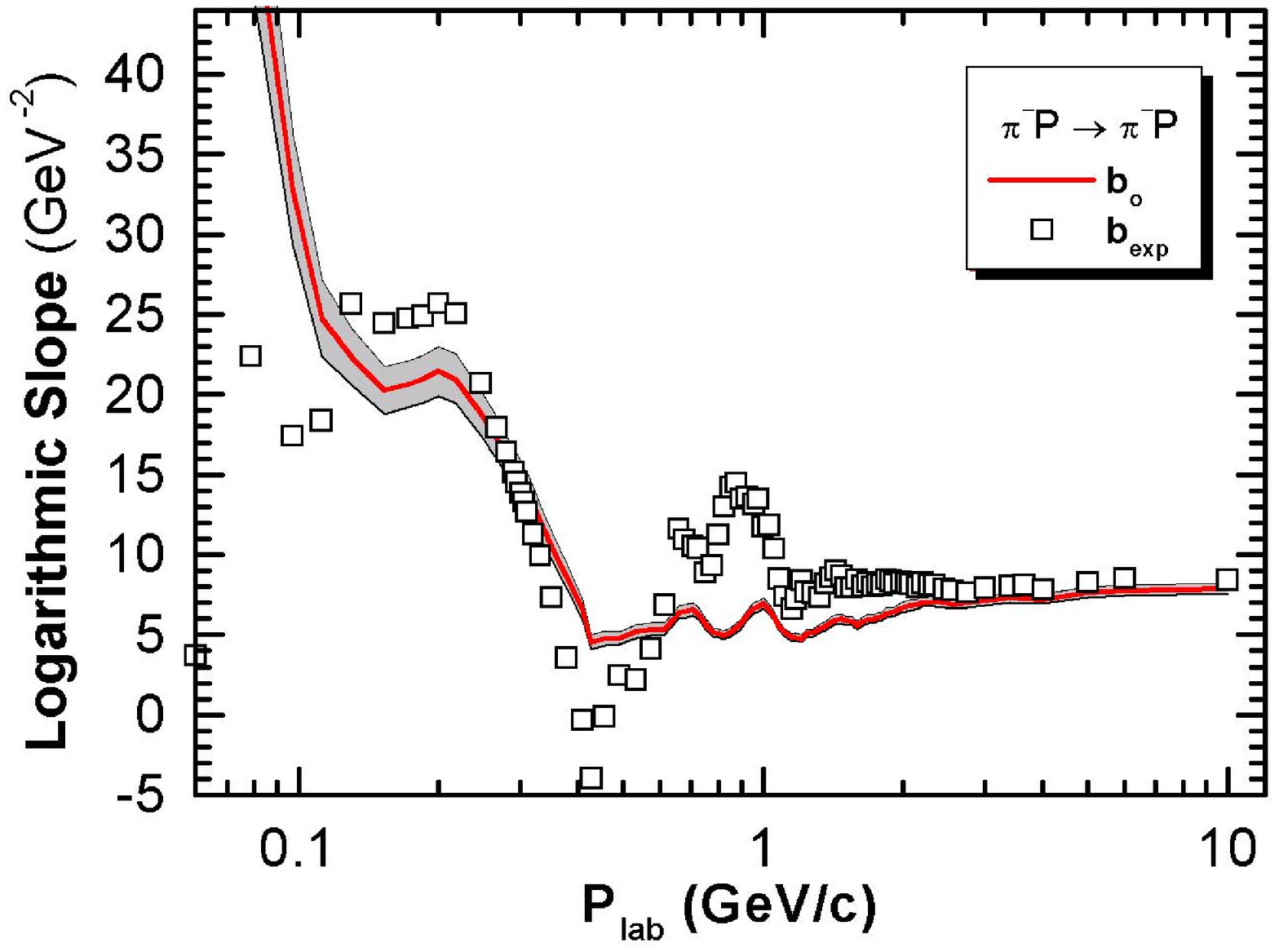}}
\caption{The experimental logarithmic slopes ($b_{exp})$ of the
diffraction peak, for the forward $\pi ^ + P \to \pi ^ + P$
scattering, are compared with the PMD-SQS-optimal predictions
$b_{o}$ (13) (see the text and Table 2).} \label{fig2}
\end{figure}

\end{document}